\documentclass[aps,prb,twocolumn,superscriptaddress]{revtex4-1}

\usepackage{xcolor}
\usepackage{soul}
\usepackage{enumitem}
\usepackage{multirow}
\usepackage{amsmath}
\usepackage{graphicx}
\usepackage{color}
\usepackage{combelow}

\newcommand{\out}[1]{}

\begin{document}

\title{Electrode effects on the observability of destructive quantum interference in single-molecule junctions}

\author{O. Sengul}
\affiliation{Institute for Theoretical Physics, Vienna University of Technology, 1040 Vienna, Austria}
\author{Angelo Valli}
\affiliation{Institute for Theoretical Physics, Vienna University of Technology, 1040 Vienna, Austria}
\author{Robert Stadler}
\affiliation{Institute for Theoretical Physics, Vienna University of Technology, 1040 Vienna, Austria}

\begin{abstract} 
Destructive quantum interference (QI) 
has been a source of interest as a new paradigm for molecular electronics 
as the electronic conductance is widely dependent on the occurrence or absence of destructive QI effects. 
In order to interpret experimentally observed transmission features, 
it is necessary to understand the effects of all components of the junction on electron transport. 
We perform non-equilibrium Green's function calculations 
within the framework of density functional theory to assess the structure-function relationship 
of transport through pyrene molecular junctions with distinct QI properties. 
The chemical nature of the anchor groups and the electrodes 
controls the Fermi level alignment, which determines the observability of destructive QI.
A thorough analysis allows to disentangle the transmission features arising 
from the molecule and the electrodes. 
Interestingly, graphene electrodes introduce features in the low-bias regime, 
which can either mask or be misinterpreted as QI effects, 
while instead originating from the topological properties of the edges. 
Thus, this first principles analysis provides clear indications to guide the interpretation 
of experimental studies, which cannot be obtained from simple H\"uckel model calculations.
\end{abstract}

\maketitle

\section*{Introduction}

In recent years it has been established, both theoretically and experimentally, 
that the electron transport properties in single-molecule junctions 
are dominated by quantum mechanical effects. 
In particular, as the propagation of electrons can be described in terms of de Broglie waves, 
the electronic transmission function exhibit quantum interference (QI) 
which results in constructive (CQI) or destructive (DQI) interference patterns.~\cite{lambert2014basic} 
While CQI results in an enhancement of the electron transmission probability, 
the hallmark of destructive QI is the presence of antiresonances in the electronic transmission function 
which can significantly reduce the conductance. 
Such QI effects are common in some classes of $\pi$-conjugated systems,
which are characterized by delocalized electronic states.  

As QI effects can be tailored by chemical~\cite{markussen2011graphical,guedon2012observation} 
and physical~\cite{frisenda2016mechanically,caneva2018mechanically,stefani2018large,sowa2018spiro,li2019symmetry,li2019gate} mechanisms and can survive even at room temperature,~\cite{wang2020scale,markussen2011graphical} 
they are relevant for the realization of functional nanoelectronic applications, 
such as memory cells,~\cite{stadler2003modulation} logic gates,~\cite{stadler2004integrating} single-molecule transistors,~\cite{stafford2007quantum,jia2018quantum} 
molecular switches,~\cite{markussen2010electrochemical,daaoub2020switching,greenwald2020highly} 
and spin filters.~\cite{saraiva-souza2014molecular,valli2018quantum,valli2019interplay} 
Particularly interesting is the potential role of QI in thermoelectric devices,~\cite{stadler2011controlling,markussen2013phonon,famili2017suppression,klockner2017tuning,miao2018influence,sadeghi2019quantum,sangtarash2018connectivity} 
where both CQI and DQI have been proposed to improve device efficiency, 
as the former can directly enhances the electrical conductivity~\cite{wang2020scale} 
whereas the presence of a sharp DQI antiresonance can affect 
the Seebeck coefficient.~\cite{strange2015interference,kloecker2017thermal,miao2018influence,grace2020connectivity}

Moreover, QI phenomena are not limited to single molecules, but have also been observed 
in other $\pi$-conjugated systems, including a variety of graphene nanostructures, 
such as nanoconstrictions,~\cite{gehring2016quantum} nanoribbons,~\cite{rosales2009conductance,calogero2019quantum}  nanoflakes,~\cite{valli2018quantum,valli2019interplay,nictua2020robust} and carbon nanotubes,~\cite{tsuji2007large} 
as well as metal-organic complexes, such as porphyrin~\cite{nozaki2015switchable,richert2017constructive}
and ferrocene~\cite{zhao2017quantum,zhao2019dft,camarasa2020mechanically} molecules, 
and in self-assembled monolayers of aromatic molecular cores.~\cite{wang2020scale} 
The possibility to exploit QI effects in molecular junctions with graphene functional blocks 
is highly desirable to pave the path towards an environment-friendly nanotechnology 
with biodegradable and inexpensive materials. 
Graphene has already been employed as a protective layer for metallic electrodes~\cite{hueser2015electron} 
and to realize asymmetric junctions.~\cite{zhang2016graphene,zhang2018technical,tao2019graphene,he2020charge} 
Another advantage of graphene is that present technologies allow the realization of atomically precise 
junctions,~\cite{cai2010atomically,chen2015molecular,caneva2018mechanically,elabbassi2019robust,caiyao2020fabrication} 
which, at the same time, offers mechanical stability beyond room temperature~\cite{prins2011room} 
and a higher degree of experimental reproducibility, 
which is crucial 
in order to achieve a precise control over QI phenomena.
However, it is also known that graphene electrodes introduce additional transmission channels, 
which can dominate in the low-bias regime and potentially obscure 
molecule-intrinsic features.~\cite{ryndyk2012edge,gehring2017distinguish} 
In this work we carry out a systematic analysis which allows us to 
disentangle the features arising from the molecule and the electrodes. 

Hitherto, a clear relation between the structure and the QI properties of a junction remains elusive. 
For this reason, the ability to predict the occurrence of an antiresonance 
and to understand its dependence on the chemical composition of the molecular junction 
plays a pivotal role for developing novel technologies based on QI. 
A few techniques, based on simplified model assumptions, allow to predict QI effects on the basis of  graphical,~\cite{markussen2010relation,stuyver2015back,oextended}
diagrammatic,~\cite{pedersen2015illusory} 
topological,~\cite{geng2015magic,tsuji2016close,tsuji2018quantum} or 
symmetry~\cite{zhao2017quantum,valli2018quantum,valli2019interplay} considerations, 
without the need of numerically expensive first-principles electron transport calculations. 
While the value of the predictive power of such {\it back of the envelope} techniques is undeniable, 
they are nevertheless unable to capture the detailed effects of the chemical nature 
of all junction components, including the leads and the linker groups, on the QI features. 

In an attempt to bridge the worlds of molecular junctions and graphene with this perspective in mind, 
we present a systematic analysis of the impact of different anchor groups and electrodes on the QI features. 
To this end, we consider molecule junctions with a pyrene core, 
which produces specific and predictable QI patterns, 
and due to its nature as a polycyclic aromatic hydrocarbon, 
can also be considered as a minimal graphene-like functional block for nanoelectronics.~\cite{sangtarash2015searching} 
Our starting point is a prototypical setup in which the pyrene subunit is attached to Au electrodes  
with thiol anchoring groups. We explore the effect of subsequently changing the anchoring groups 
and the electrodes to build a hydrocarbon junction, 
which is relevant for understanding the electron transport through graphene nanoconstrictions. 
Specifically, we substitute thiol- with acetylene-terminated linkers and analyze 
the changes in the transport properties due to the direct Au-C bonds 
and the charge redistribution within the junction. 
Next, we substitute the Au electrodes with graphene nanoribbons (GNRs). 
While the featureless density of states of Au electrodes close to the Fermi level 
have little effect on the shape of the transmisson function, 
the electronic properties of graphene nanostructures are strongly affected by the topology of their edges. 
We shall see that this has profound consequences for the electron transport through the junction,  
in particular concerning the observability of DQI. 
By considering graphene electrodes with different size, 
we are able to disentangle unambiguously the transport features that are intrinsic to the molecule 
from those stemming from the chemical and topological nature of the electrodes.

\begin{figure}[t]
    \begin{center}
        \includegraphics[width=0.48\textwidth]{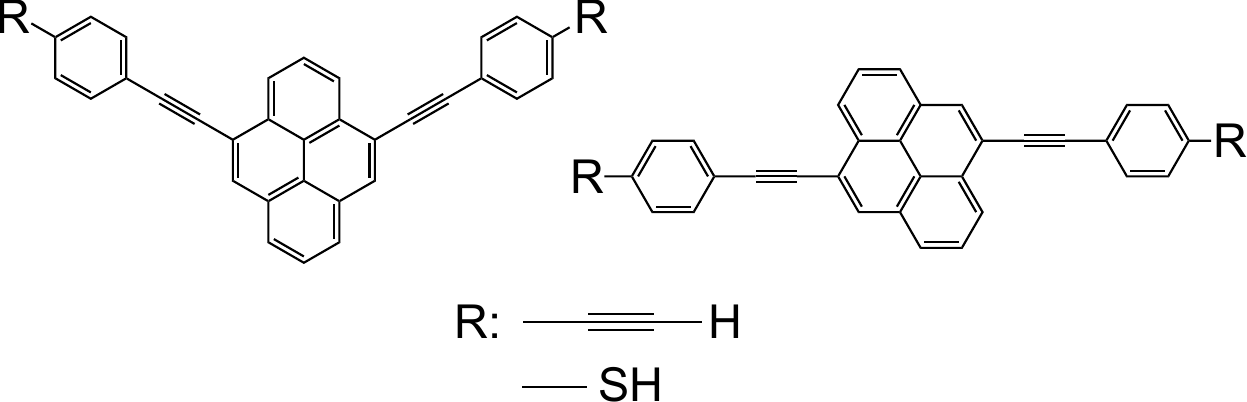}
    \end{center}
    \caption{Molecular structures of the meta- and para-substituted pyrene with propynylbenzene linkers. 
    R denotes the anchoring groups, being either thiol or acetylene. }
    \label{Structures}
\end{figure}

\section*{QI in molecular junctions with pyrene core}\label{sec:QI-pyrene}

In the following, we consider molecular junctions where a pyrene core  
is connected to the electrodes in two different configurations via propynylbenzene linkers 
with thiol- and acetylene-terminated anchor groups, as shown schematically in Fig.~\ref{Structures}. 

In a realistic scenario, one should expect that the transport properties of a molecular junction 
are determined by the nature of the linkers and the anchor groups, as well as the electrodes, 
besides the position of the contact atoms on the $\pi$-conjugated molecule itself. 
In general, it is difficult to establish a clear relation between the structure of its components 
and the QI properties of a molecular junction.
Graphical methods~\cite{stadler2004integrating,markussen2010relation,pedersen2015illusory} 
allow, to some extent and on qualitative grounds, to predict the occurrence of QI antiresonances, 
but become hardly applicable with in complex systems. 
For this reason our analysis of QI features of pyrene molecular junctions is twofold. 
First, we focus on the pyrene core and we employ a graphical scheme 
and calculate the transmission function within a simple H\"uckel model (SHM),  
to predict and demonstrate that the aromatic molecular subunit is the source of DQI. 
For a quantitative analysis we rely on density functional theory (DFT) calculations 
in combination with non-equilibrium Green's function (NEGF) formalism, 
which allows us not only to verify the qualitative predictions, 
but also to investigate the quantitative effect of different anchor groups and electrodes 
on the QI features.

\begin{figure}[b!]
    \centering
    \includegraphics[width=0.48\textwidth]{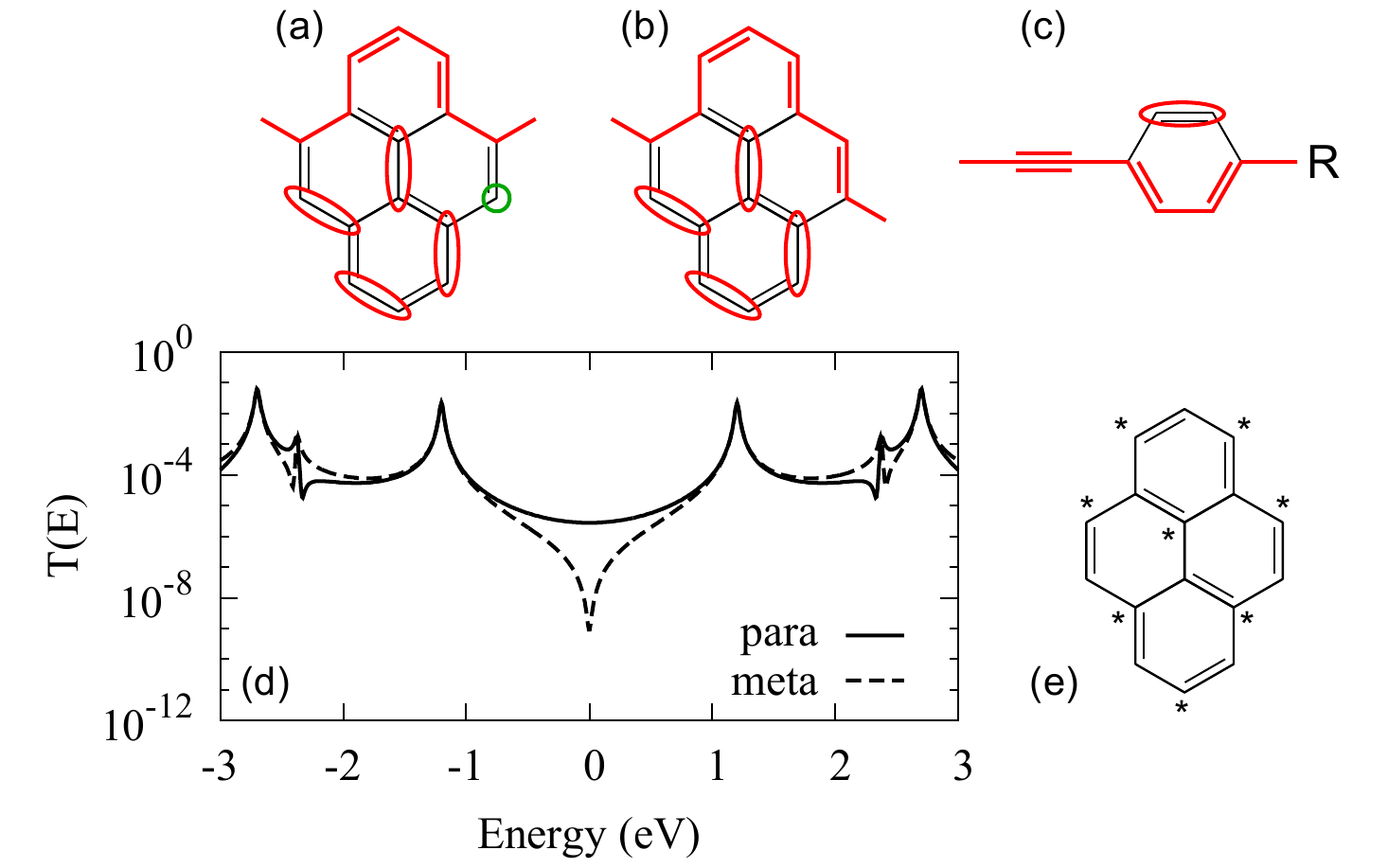}
    \caption{Analysis of QI properties of the pyrene core in the meta- (a) and para- (b) configurations, 
    and of the linker (c) with the graphical scheme. 
    The linker does not introduce sources of DQI, independently of the anchor group R. 
    (d) Transmission function of the pyrene junction within the H\"uckel model 
    with a wide-band limit approximation for the leads. 
    (e) As an alternant hydrocarbon, pyrene can be divided into starred and unstarred sublattices. }
    \label{fig:pyrene-QI}
\end{figure}

We apply the graphical scheme~\cite{stadler2004integrating,markussen2010relation} 
to the pyrene core connected in the meta- and para-configurations. 
The scheme prescribes to draw a continuous path through neighboring atoms 
from the left to the right lead. 
No DQI is predicted if such a path 
crosses all sites of the molecule, 
or if the remaining sites can be grouped into either pairs or closed loops (red ellipses). 
Representative paths for the meta- and para-configurations
are shown in  Figs.~\ref{fig:pyrene-QI}(a,b), respectively. 
It is easy to convince oneself that any possible path in the meta configuration leave 
at least an unpaired site (green circle). 
Hence, we predict the occurrence of a QI antiresonance in the meta configuration. 
Including the linkers (with either thiol- or acetylene-termination) in the graphical scheme 
do not change 
this conclusion, 
since linear chains and para-connected benzene rings do not result in DQI, 
see Fig.~\ref{fig:pyrene-QI}(c) for the corresponding graphical analysis.~\cite{markussen2010relation} 

The graphical predictions can be easily verified by considering a SHM for the pyrene junction. 
We set the energy of the C~2$\text{p}_z$ atomic orbitals (AOs) to zero 
and assume a overlap between neighboring AOs of $\beta=2.7$~eV, 
which is a typical value for graphene,~\cite{castro-neto2009graphene} 
so that the Hamiltonian of the SHM is proportional to pyrene's adjacency matrix as 
$[{ H}_{\text{SHM}}]_{ij}=\beta{ A}_{ij}$ 
with $A_{ij}=1$ for AOs $i$ and $j$ sharing a bond, and zero otherwise.~\cite{tsuji2018quantum}
The retarded Green's function of the molecule reads

\begin{equation}
 { G}(E) = \big[ (E+\imath\eta){ I} - { H}_{\textrm{SHM}} + { \Sigma}^L + { \Sigma}^R \big]^{-1},
\end{equation} 
and the Landauer transmission function~\cite{landauerJRD1} is given by   
\begin{equation} 
\label{eq:Te_landauer}
 T(E) = \mathrm{Tr} \big[ { \Gamma}^L(E) { G}^{\dagger}(E) { \Gamma}^R(E) { G}(E) \big], 
\end{equation} 
where ${ \Gamma}=\imath({ \Sigma}^{\phantom{\dagger}}-{ \Sigma}^{\dagger})/2$
denotes the coupling to either the left (L) or right (R) lead 
in terms of the corresponding embedding self-energy. 
As pyrene is contacted to the leads through a single H\"uckel AO, 
each matrix ${ \Gamma}$ has a single non-zero matrix element, 
$\Gamma^L_{\ell \ell}$ and $\Gamma^R_{rr}$, respectively, 
which we assume to be energy independent (wide-band limit approximation). 
The transmission function hence reduces to  
\begin{equation} \label{eq:Te_single}
 T(E) = \Gamma^L_{\ell \ell}\Gamma^R_{rr} |G_{\ell r}(E)|^2  
\end{equation} 
and displays resonances in correspondence to the energy H\"uckel molecular orbitals (MOs) of pyrene. 
In the meta configuration, the transmission function also 
exhibits an antiresonance in the middle of the HOMO-LUMO gap (i.e., at $E=0$). 
The antiresonance orginates from DQI, in complete agreement with the scenario 
predicted within the graphical method. 

Analogous conclusions can be drawn from the Coulson-Rushbrooke pairing theorem,~\cite{zhao2017destructive} 
which states that for alternant hydrocarbons, when the contact sites $\ell$ and $r$ belong to the same sublattice, 
the contribution to the Green's function coming from the MOs cancels pairwise. 
Each pair $k$ is formed by an occupied and an unoccupied MO.   
Then, the spectral representation of the Green's function in the MO basis reads 
\begin{equation} \label{eq:GlrMOs}
 G_{\ell r}(E) = \sum_{k=1}^{N/2} \Big[ 
                          \frac{c^{\phantom{\ast}}_{\ell;-k}c^{*}_{r;-k}}{E+\imath \eta - \epsilon_{-k}} +
                          \frac{c^{\phantom{\ast}}_{\ell: k}c^{*}_{r; k}}{E+\imath \eta - \epsilon_{k}}
                          \Big], 
\end{equation}
where $c_{i;k}$ is the coefficient of MO $k$ at site $i$ and $N$ the number of MOs in the Hamiltonian. 
The pairing theorem predicts that for alternant hydrocarbons each pair fulfills 
the conditions $\epsilon_{k}=-\epsilon_{-k}$ and that all coefficient in one sublattice have opposite sign 
between the occupied and unoccupied MOs, 
then for each $k$ the contribution of the pair to the sum in Eq.~(\ref{eq:GlrMOs}) 
cancels exactly at $E=0$.~\cite{zhao2017destructive}
In the meta configuration, $\ell$ and $r$ belong to the same sublattice of the pyrene molecule, 
as shown in Fig.~\ref{fig:pyrene-QI}(e), 
and the Coulson-Rushbrooke predicts a QI antiresonance at the Fermi level, 
in agreement with the graphical scheme and the SHM calculation.

\section*{DFT+NEGF Electron transport calculations} \label{sec:DFT+NEGF}

The above analysis establishes clearly the occurrence of DQI 
depends on the position of the contact atoms on the pyrene core. 
While this scenario is qualitatively robust, we shall see in the following that 
the details of the chemical bonding between the molecule and the electrodes, 
as well as the physical properties of the electrodes, which can be addressed within DFT,
can drastically affect the observability of DQI in the transmission function.

\subsection*{Computational details} \label{sec:details}

We investigate the transport properties of molecular junctions with pyrene core 
within the DFT+NEGF formalism,~\cite{brandbyge2002DFTNE,xue2002first} 
as implemented within the Atomic Simulation Environment~\cite{larsen2017ASE} (ASE) 
and the GPAW software package.\cite{mortensen2005GPAW,enkovaara2010GPAW} 
The transmission function is calculated according to Eq.~(\ref{eq:Te_landauer}) 
where the Green's function is evaluated on the whole DFT basis set for the scattering region. 
The coupling matrix $\Gamma$ is obtained from the surface Green's function of the leads 
and the couplings between the basis function of the lead and scattering regions. 
The transport Hamiltonian of each structure is evaluated  
in a local combination of atomic orbitals (LCAO)~\cite{larsen2009LCAO} double-$\zeta$ polarized basis, 
with a Perdew-Burke-Ernzerhof (PBE) parameterization for the exchange-correlation functional, 
and a sampling of 0.2~\AA \ grid spacing. 
For each molecular configuration we perform an atomic optimization of the molecule in vacuum 
until the value of the Hellmann–Feynman forces are below $0.05$~eV/\AA.

\begin{figure*}[t!]
    \begin{center}
        \includegraphics[width=1.00\textwidth]{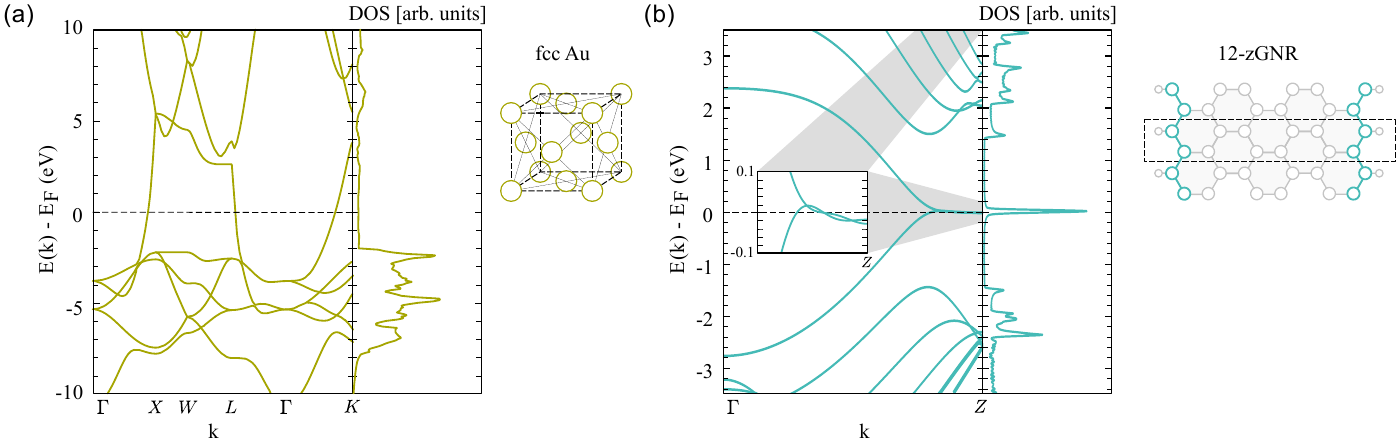}
    \end{center}
    \caption{Bandstructure $E(k)$ and density of states (DOS) of (a) fcc Au, and (b) 12-zGNRs. 
    The width of the GNR is defined as the number of C atoms along the dimer line. 
    The insets next to each panel show the corresponding unit cell (dashed lines). 
    Close to the Fermi level, Au displays a featureless DOS, 
    while zGNRs display non-dispersive bands and a corresponding peak in the DOS, 
    due to the localized states near the ZZ edges (highlighted in color in the inset). }
    \label{fig:EkDOS}
\end{figure*}

In the case of Au electrodes, the scattering region consists of seven $6\times 4$ layers of Au(111). 
The linkers are connected via anchoring groups to Au adatoms placed 
in the hollow position of the Au(111) surface. 
We take a typical bonding distances of $d_{\textrm{Au-S}}=2.12$~\AA \ for 
thiol-terminated linkers~\cite{zhao2017destructive,stadler2005forces} 
and $d_{\textrm{Au-C}}=1.92$~\AA \ for acetylene-terminated linkers.~\cite{liu2013probing} 
The scattering region is sampled with a $4 \times 4 \times 1$ Monkhorst-Pack mesh,
and the leads are sampled with a $4 \times 4 \times 6$ Monkhorst-Pack mesh, 
where $z$ denotes the transport direction. 

In the case of graphene electrodes, GNRs typically have either a zigzag (ZZ) or an armchair (AC) termination. 
We consider the GNRs with ZZ edges and denote periodic structures with width N as N-zGNRs, 
following the standard nomenclature in the literature. 
The dangling bonds at the graphene edges are passivated with H atoms. 
For the portion of GNRs included in the scattering region, we assume a rectangular termination, 
which can be expected for atomically-precise GNRs obtained 
though {\it bottom-up} synthesis by polymerization.~\cite{cai2010atomically} 
Hence, in this configuration, at the junction interface between the GNRs and the molecular bridge, 
zGNRs have a transverse edge with AC topology 
(on the contrary, aGNRS have a transverse edge with ZZ topology). 
For graphene electrodes, a natural choice is to connect the molecule to the leads 
by replacing one of the H atoms passivating the edge on each side 
with the acetylene-terminated linkers,  
with a bonding distance $d_{\textrm{C-C}}=1.54$~\AA. 
The scattering region is sampled with a $1 \times 1 \times 1$ Monkhorst-Pack mesh, 
and the leads consist of GNR unit cells sampled with a $6 \times 1 \times 1$  Monkhorst-Pack mesh, 
where $x$ denotes the transport direction.  
For infinitely-wide GNRs (semi-infinite graphene sheet) we impose periodic boundary conditions 
along the in-plane direction ($y$) transverse to the transport direction, 
which we sample with a Monkhorst-Pack mesh with up to $256$ $k$-points.

\section*{Results and Discussion} \label{sec:results}

From the atomic and electronic structure perspective, Au and graphene are completely different materials.
While for Au electrodes the transmission close to the Fermi level 
is dominated by a single transmission channel 
with predominant Au~6$\text{s}$ character,~\cite{Cuevas2010Molec-12488,jia2013review} 
the electronic properties of graphene critically depend on the topology of the edges. 
In particular, it is well known that GNRs with ZZ termination 
support non-dispersive states localized near the edges. 
Hence, the zGNRs that we consider in the following are metallic, 
while aGNRs are typically semiconductors.~\cite{wakabayashi1999electronic,son2006energy} 
Moreover, aGNR electrodes have transverse ZZ edge states 
that decay very slowly into the bulk,~\cite{li2014anomalous,li2015lowbias}  
making the correct computational treatment with first principle methods very challenging. 
To illustrate the properties of the electrodes,  
in Fig.~\ref{fig:EkDOS} we compare the bandstructure $E(k)$ and corresponding density of states (DOS) 
of fcc Au, to those of periodic zGNR. 

Another difference between Au and graphene electrodes is that GNRs with ZZ edges 
have a tendency towards magnetic ordering. 
Hence, in order to complete our analysis, we also discuss the effect of a spin-symmetry breaking 
on the DQI properties of pyrene molecular junction.

\subsection*{Au electrodes} \label{sec:Au}

\begin{figure*}[htbp]
    \centering
    \includegraphics[width=1.00\textwidth]{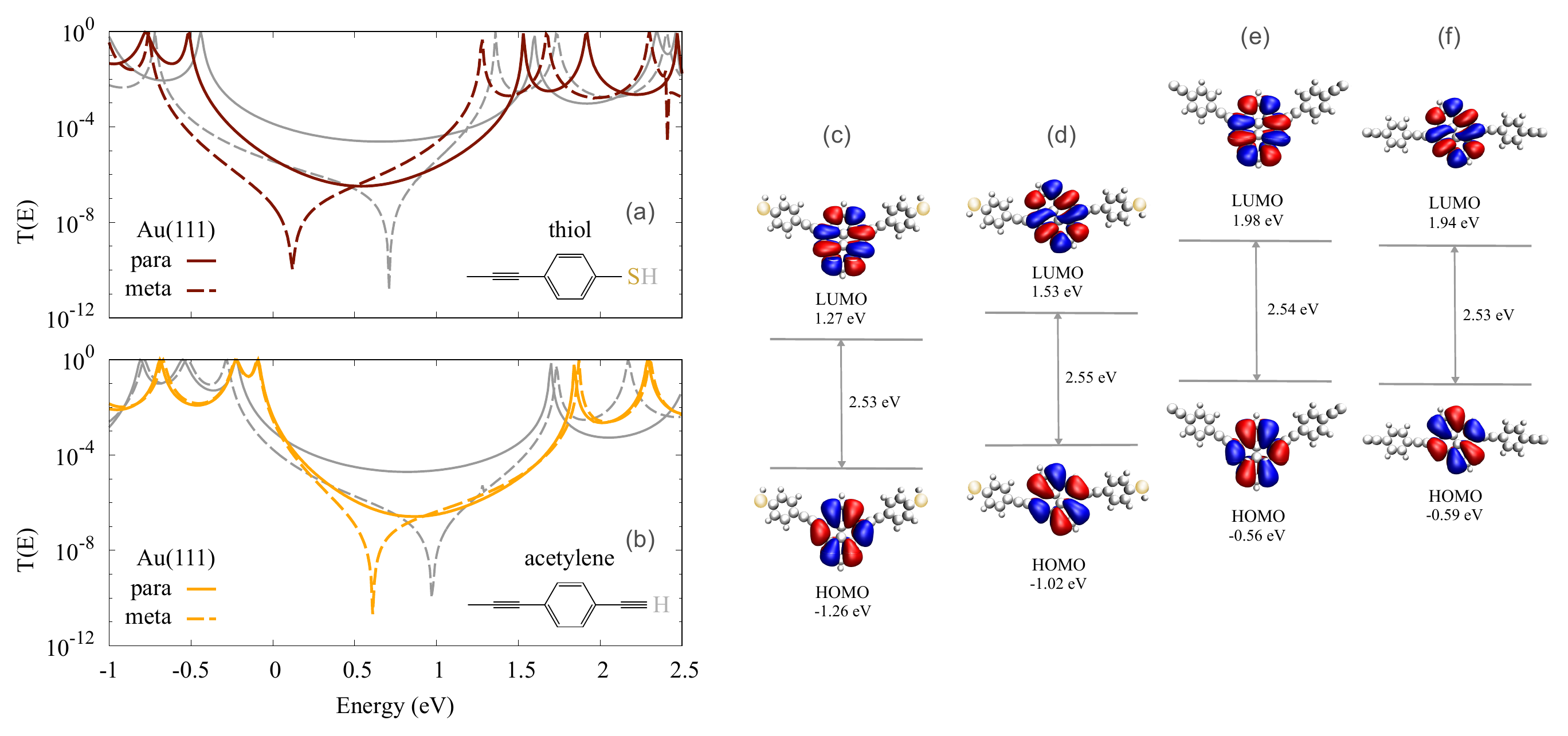}
    \caption{(a,b) Transmission functions in the meta- and para-configurations 
    connected with different anchor groups to Au electrodes. 
    The grey lines represent the transmission function obtained 
    with only the two pairs of MOs closest to the Fermi level. 
    The insets show the linkers connecting the pyrene molecule to the Au leads. 
    (c-f) Frontier MOs obtained by subdiagonalization of the pyrene core and corresponding HOMO-LUMO gaps. }
    \label{fig:Au}
\end{figure*}

We first discuss the case in which the molecule bridges Au electrodes. 
The transmission as a function of energy of the incoming electrons 
for the pyrene connected in the meta- and para-configurations 
with either thiol or acetylene anchoring groups is shown in Fig.~\ref{fig:Au}(a,b), respectively.  
The transmission function is similar in all cases, 
with the main difference being the antiresonance within the HOMO and LUMO gap
present in the meta configuration. 
This is in complete agreement with the previous analysis with the graphical scheme, 
the Coulson-Rushbrooke pairing theorem, and the SHM. 
However, the conductance of the junction also strongly depends on the relative positions 
of the HOMO, the LUMO, and the antiresonance with respect to the Fermi level, 
which we are going to analyze in detail in the following. 

In all cases, the antiresonance is closer to the HOMO than to the LUMO. 
For thiol-terminated linkers, $\epsilon^{\textrm{DQI}}\approx0.12$~eV 
is also close to the Fermi level $E_F=0$, 
whereas for acetylene-terminated linkers there is a shift to higher energies, 
resulting in $\epsilon^{\textrm{DQI}}\approx0.6$~eV. 
The most striking difference between the two cases is the relative position of the Fermi level 
within the HOMO-LUMO gap. 
In order to shed light on this, we perform a Bader charge analysis~\cite{tang2009grid} 
for the free molecule and the junction with both linkers, 
because the Fermi level alignment is closely related to the 
zero-bias charge transfer.~\cite{stadler2006fermi,stadler2007fermi,stadler2010conformation,kastlunger2013charge} 
The HOMO-dominated profile in the case of acetylene linkers is coming from 
the negatively charged molecule after the placement in between the leads, 
which shifts the position of HOMO and LUMO to higher energies, 
with respect to thiol linkers.~\cite{stadler2006fermi,stadler2007fermi} 
When we analyze the case with thiol linkers, there is a clear difference in the Fermi level alignment 
of the meta- and para-configurations. This can be explained by considering the larger 
electron transfer from the molecule towards the electrodes in the meta 
with respect to the para configuration 
as reported in Table~\ref{table:Au}. 
In the case of acetylene linkers instead, the charge transfer is negative, corresponding to electron transfer from the electrodes to the molecule. However, there is no direct relation between $\epsilon_{\textrm{HOMO}}$ and $\Delta Q$ 
and the molecular level position cannot be fully explained by net charge transfer alone, 
but the molecular dipole moments, can possibly account for it. 
The dipole moment is larger in the meta configuration and it is orientated within the plane of the molecule 
and orthogonal to the transport direction (see Table~\ref{table:Au}). 

\begin{table}[b]
\centering
\small
  \caption{\ Position of the HOMO resonance, obtained from the subdiagonalization of the molecule including the linkers, net charge on the molecule $\Delta{Q}$, dipole moment of the free molecule $\mu$ in all configuration. }
  \label{table:Au}
  \begin{tabular*}{0.45\textwidth}{@{\extracolsep{\fill}}ccccc}
    \hline
     & & $\epsilon_{\text{HOMO}}$ (eV) & $\Delta{Q}$ ($|e|$) & $\mu$ (${\textit{D}}$) \\
    \hline
    Thiol     & meta & -0.97 & \phantom{-}0.39 & 0.33 \\
              & para & -0.72 & \phantom{-}0.16 & 0.19 \\
    Acetylene & meta & -0.20 &           -0.38 & 0.14 \\
              & para & -0.20 &           -0.50 & 0.01 \\
    \hline
  \end{tabular*}
\end{table}

It is possible to qualitatively describe the QI effects 
taking into account only the symmetry properties of the pyrene core. 
This analysis is also consistent with our DFT calculations. 
We diagonalize the sub-block of the transport Hamiltonian of the pyrene core  
to obtain the projected MOs. 
We can calculate an approximate transmission function selecting the two pairs of MOs 
closest to the Fermi level, 
i.e., $\textrm{HOMO}-1$ , $\textrm{HOMO}$, $\textrm{LUMO}$, and $\textrm{LUMO}+1$ 
and cutting out the rest.~\cite{stadler2009quantum,guedon2012observation,zhao2017destructive,gandus2020smart} 
As shown in Fig.~\ref{fig:Au}(a,b) with grey lines, the contribution of these two pairs is sufficient 
to qualitatively reproduce the antiresonance. 
The approximate transmission function shows a good qualitative agreement with the one 
from the original transport Hamiltonian 
but the position of the antiresonance and the Fermi level alignment  
are more sensitive to such simplifications.

The symmetry of the MOs carries relevant information on QI. 
According to the Coulson-Rushbrook pairing theorem, 
the antiresonance is a consequence of pairwise cancellation 
of the contributions of the MOs to the transmission. 
The frontier MOs of the pyrene core are shown in Figs.~\ref{fig:Au}(c-f) 
for thiol- and acetylene-anchoring group in the meta- and para-configurations. 
In the case of thiol anchoring groups, 
the signs of the amplitudes of the eigenvector at the positions where pyrene is connected to the linkers 
are opposite for both the HOMO and the LUMO in the meta configuration. 
Hence, we expect a cancellation of the contribution of the frontier MOs in the transmission function, 
resulting in an antiresonance at some energy $\epsilon^{\textrm{DQI}}$, 
not necessarily at the Fermi level, or mid-gap as in the SHM calculation, 
since the spectrum is not particle-hole symmetric.  
Instead, in the para configuration, the amplitude of the LUMO has the same sign at the contact sites, 
and therefore there is no cancellation in the transmission. 
When we substitute thiol- with acetylene-group as anchoring units, 
the same symmetry considerations hold. 
This suggests that while the origin of the QI phenomena does not depend on the nature 
of the linkers and anchoring groups, the energy at which the contributions from all MOs cancels 
is related in a non-trivial way to the chemical nature and the electronic structure 
properties of all the components of the junction.

\subsection*{Graphene electrodes} \label{sec:graphene}

\begin{figure}[htbp]
    \centering
    \includegraphics[width=0.48\textwidth]{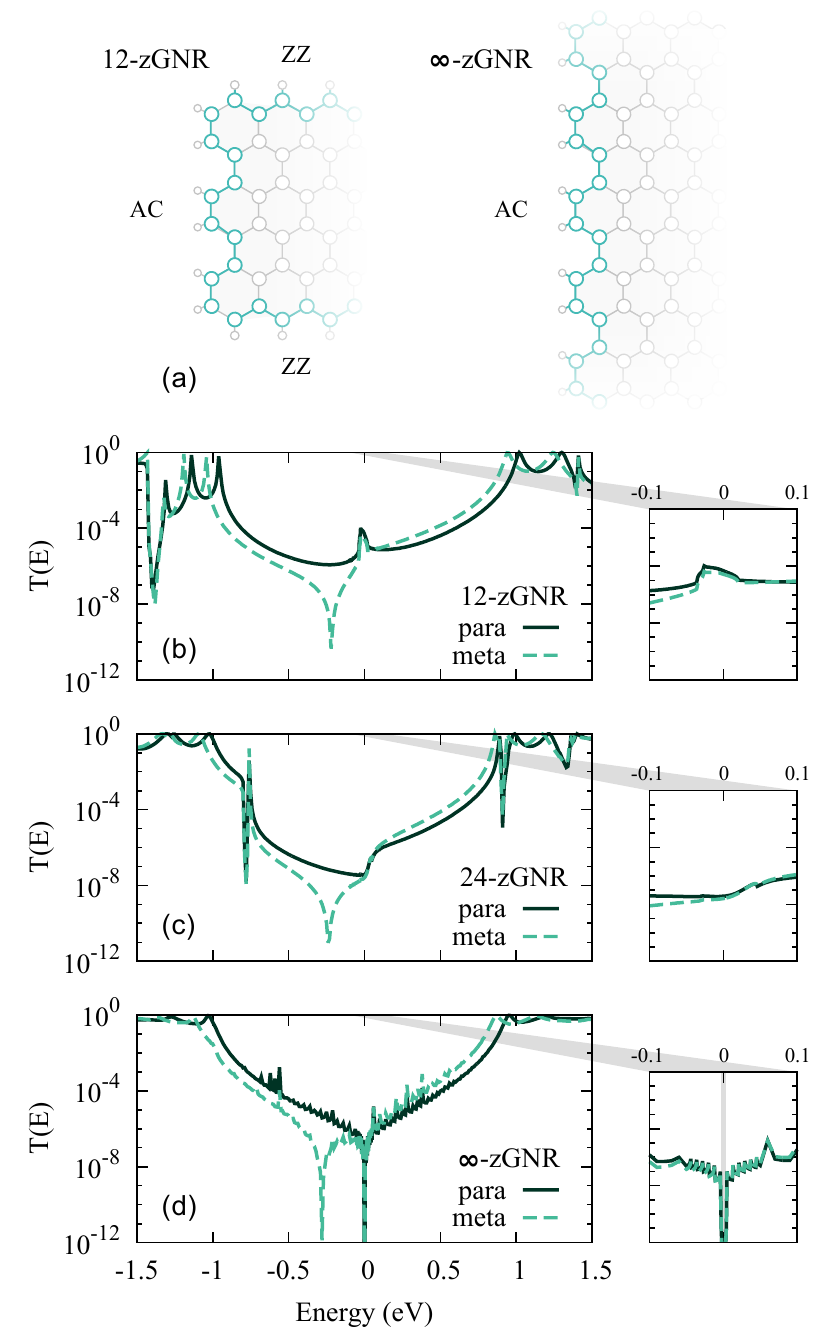}
    \caption{Transmission functions for pyrene junctions in meta- and para-configurations 
    with zGNR electrodes of different width. 
    The width of the zGNR is defined as the number of C atoms along the dimer line. 
    Each inset highlights the low-energy features of the transmission function close to the Fermi level. }
    \label{fig:GNR_Te}
\end{figure}

We examine now the case in which the pyrene molecule is connected to graphene electrodes 
via acetylene-terminated linkers. 
The schematic representations of the junctions with GNR electrodes are shown in Fig.~\ref{fig:GNR_Te}(a) 
where the edges transverse and parallel to the transport direction are highlighted in color. 
As anticipated, 
zGNRs have AC edges transverse to the transport direction.  
In order to assess the effect of edge size and topology on DQI, 
we consider zGNRs with different widths.

Since  with Au electrodes the QI properties arise from the position of the contact atoms 
in the pyrene, it is reasonable to expect similar features also for graphene electrodes, 
and an antiresonance is found indeed only for the meta-connected structures. 
In contrast to the Au electrodes, in which the transmission is HOMO-dominated, 
here the Fermi level is located close to the middle of the HOMO-LUMO gap 
and the 
antiresonance is below $E_F$. 
This is not surprising considering that with GNR electrodes, 
the junction is hydro-carbon throughout and the charge distribution 
between the molecule and the leads is expected to be uniform.~\cite{alqahtani2018breakdown} 

In all transmission functions showed in Fig.~\ref{fig:GNR_Te}, 
there are features at the Fermi level, which were absent in the case of Au electrodes. 
The most natural explanation is that these features stem from 
graphene edge states. 
Indeed, the shape of the transmission reflects the presence 
of both ZZ and AC edges, in a way that depends on the width and the topology of the electrodes. 
In Figs.~\ref{fig:GNR_Te}(b), we observe a resonance at the Fermi level, 
which indicates a transmission mechanism involving the edge states of the 12-zGNRs. 
The transmission feature is significantly broad due to the high DOS of the zGNR leads. 
Some feature, albeit not a resonance is also observe in Fig.~\ref{fig:GNR_Te}(c), for the 24-zGNRs. 
We speculate the overlap between the edge state and the states of the molecular bridge 
decreases by increasing the width of the zGNR. 
Eventually, for $\infty$-zGNR, the transmission function in Fig.~\ref{fig:EkDOS}(c) 
recovers a \emph{V-shape}, reminiscent of the DOS of bulk graphene, 
but it exhibits a gap at the Fermi level (see inset) 
due to the presence of the transverse edge of the electrode, which has AC character.

%

\section*{Effect of DQI on the I-V characteristics} \label{sec:IV}

\noindent
In an experimental setup, the current-voltage (I-V) characteristics 
is measured rather than the transmission function. 
In order to draw conclusions on the observability of DQI effects, 
we evaluate the electric current (per spin) as 
\begin{equation}
    I = \frac{e}{h} \int_{\infty}^{\infty} dE \ T(E) \ \big[ f_S(E) - f_D(E) \big],
\end{equation}
where $e$ is the electric charge and $h$ the Planck constant. 
The Fermi distribution function of the source (S) and drain (D) electrodes are given by
\begin{equation}
 f_{S/D}(E) = \frac{1}{1 + \exp[(E - V_{S/D})/k_BT]}, 
\end{equation}
where $k_B$ is the Boltzmann constant and $V_b=V_{S}-V_{D}$ is the symmetric bias drop 
between source and drain. 
In the calculations, we assume the thermal broadening of the Fermi distribution 
to be $k_BT = 5$~meV and the current is evaluated neglecting the bias dependence 
of the transmission function, i.e., $T(E, V_b) \approx T(E)$. 
In the following, we also look at the differential conductance $G=dI/dV_b$ as a function 
of both bias ($V_b$) and gate ($V_g$) voltages. 

\begin{figure}[htbp]
    \centering
    \includegraphics[width=0.465\textwidth]{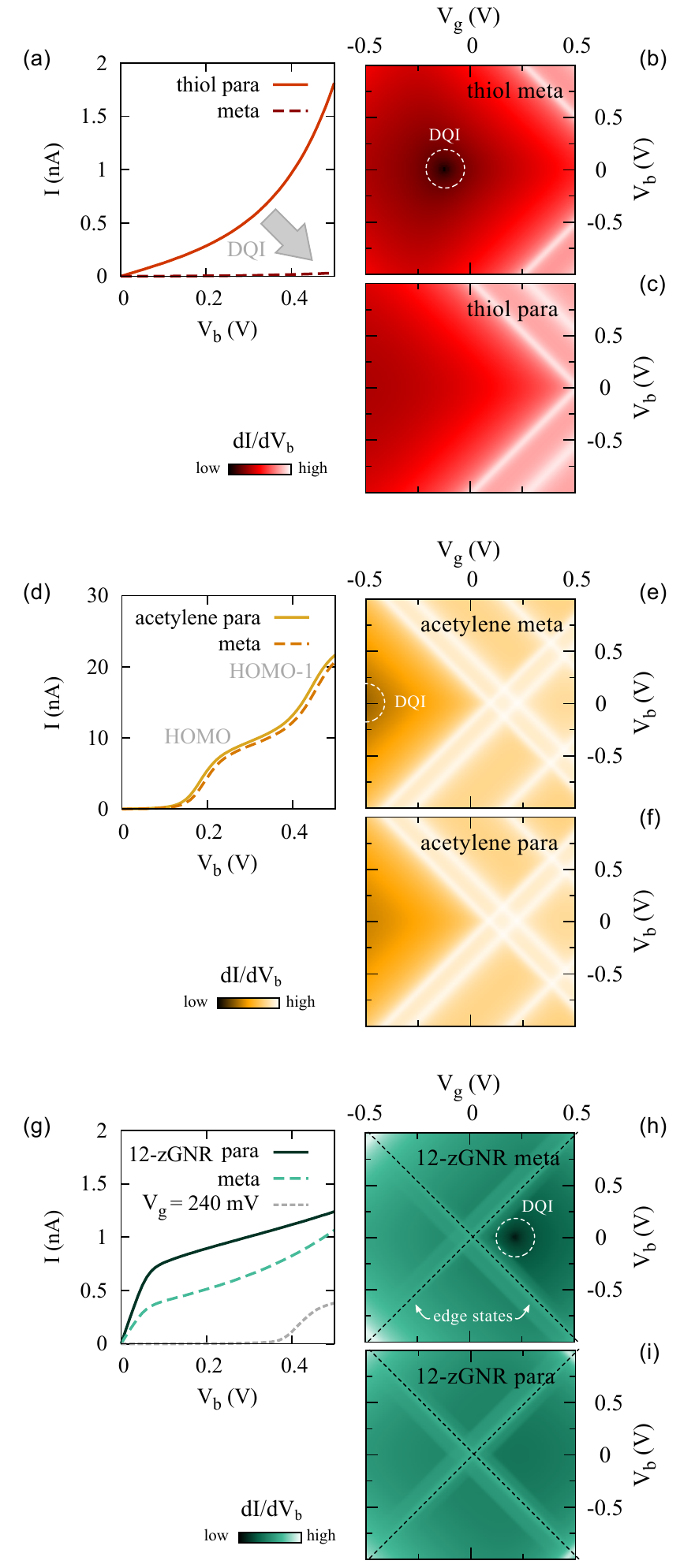}
    \caption{I-V characteristics (a,d,g) and differential conductance $G=dI/dV_b$ as a function 
    of gate ($V_g$) and bias ($V_b$) voltages (b,c,e,f,h,i) for all junctions. 
    For Au electrodes, DQI suppresses electron transport in the meta configuration 
    in the case of thiol anchoring groups, 
    while for acetylene groups it is dominated by resonant tunneling though the HOMO 
    and QI has no visible effects. 
    For graphene electrodes, the effect of DQI is partially masked 
    due to the presence of the zGNR edge states but it can be enhanced 
    by applying a gate voltage $V_g=240$~mV to align the Fermi level to the antiresonance. }
    \label{fig:IV-analysis}
\end{figure}

The calculated I-V characteristics and the differential conductance maps 
for the junctions with Au and zGNR electrodes reveal the qualitative relationship 
between the transmission and the transport features. 
The results are presented in Fig.~\ref{fig:IV-analysis}. 
In the case of Au electrodes, the outcome is quite interesting. 
For all combinations of contact configurations, the HOMO-LUMO gap is very similar, 
while the junctions with acetylene linkers exhibit an overall much higher current  
compared to the thiol ones, reaching the order of a few $\mu$A, see Figs.~\ref{fig:IV-analysis}(a,d). 
This is related to the close proximity of the HOMO to the Fermi level 
and its wider peak in the transmission function for this linker type.~\cite{schwarz2014high} 
As a consequence, the meta- and para-configurations display very similar I-V characteristics 
since the effect of DQI is negligible with respect to the resonant contribution of the HOMO.  
The current displays clear steps in correspondence to the MOs entering the bias window, 
as indicated in Fig.~\ref{fig:IV-analysis}(d). 
For thiol-terminated linkers, the combination of two effects, i.e., the Fermi level 
being closer to the middle of the HOMO-LUMO gap and closer to the antiresonance, 
results in a large difference in the I-V characteristics of the meta- and para-configurations. 
In Figs.~\ref{fig:IV-analysis}(b,c), we show the map of the differential conductance, 
where the effects of DQI is clearly visible, by comparing the meta- and para-configurations 
for this combination of electrodes and anchoring groups. 
The bright lines in the I-V characteristics correspond to the resonances of the MOs. 

It is interesting to observe the effect of edge topology of the graphene electrodes on the I-V characteristics. 
The transmission features close to the Fermi level, originating from the zGNR edge states, 
appear in both meta- and para- configurations. 
As a consequence, in both cases we observe a metallic-like I-V characteristics 
and QI effects do not play an important role. 
However, for devices based on graphene, and in general 2D-materials, gating is a viable strategy 
to improve device performances. 
Specifically, in order to enhance the effects of DQI, one can introduce an external electric field 
to align the Fermi level to the antiresonance. 
Applying a gate voltage $V_g=240$~mV to the junctions with zGNR electrodes, 
results in a strongly suppressed current, see  grey dashed line in Fig.~\ref{fig:IV-analysis}(g). 
This allows one to clearly distinguish the I-V characteristics of the configuration with DQI from the other. 
In Figs.~\ref{fig:IV-analysis}(h,i) we show the map of the differential conductance. 
The resonances corresponding to the edge states are clearly visible for both meta- and para-configurations 
while suppression due to DQI is clearly identified as a single {\it dark spot} at $V_g=240$~mV. 

\section*{Effects of edge magnetism in graphene}

While the experimental evidence of magnetism in graphene nanostructures~\cite{magda2014room,slota2018magnetic} 
is still controversial, 
from the theoretical point of view it is well established that zGNRs have a tendency towards magnetic order. 
When the spin symmetry is lifted, the GNR orders according to a N\'{e}el antiferromagnetic (AF) pattern, 
which results in ferromagnetically (FM) aligned ZZ edges, due to the local sublattice imbalance. 
Opposite edges display opposite magnetization, yielding a global singlet (S=0) 
and thus fulfilling Lieb's theorem. 
The magnetic moments are rapidly suppressed towards the bulk, as well as at AC defects and corners, 
where the atoms from different sublattices are paired.
The magnetic structure of zGNRs has been widely discussed in the literature, 
and it has been shown to remain qualitatively unchanged also when 
a local electron-electron interactions is included explicitly 
within DFT+U,~\cite{kabir2014manipulation,ganguly2017magnetic} 
a mean-field Hubbard model,~\cite{fernandez2007magnetism,hu2017understanding} 
and within more sophisticated many-body techniques.~\cite{feldner2011dynamical,feldner2010magnetism,valli2016effective,valli2018quantum,baumann2020inducing,phung2020spin} 
Hence, the following discussion can be regarded as of general relevance 
for the interplay between DQI and edge magnetism.

\begin{figure}[htbp]
    \begin{center}
        \includegraphics[width=0.46\textwidth]{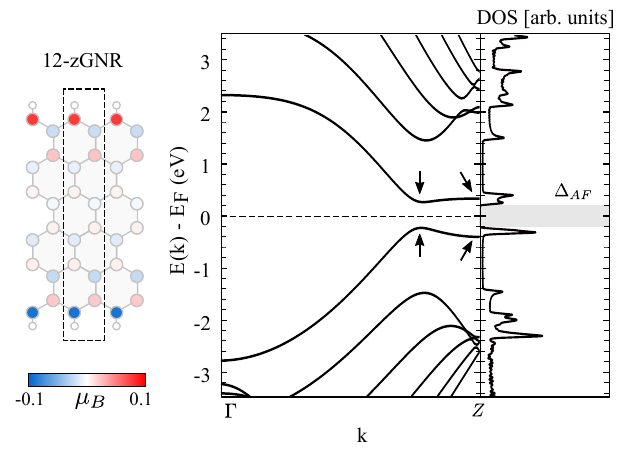}
    \end{center}
    \caption{Bandstructure $E(k)$ and density of states (DOS) of AF 12-zGNR. 
    The bandstructure of the two spin polarizations is degenerate in all bands. 
    The arrows mark the relevant energy splittings along the $\Gamma-Z$ path 
    of the one-dimensional Brillouin zone. 
    The inset in the side panel show the corresponding unit cell (dashed lines). 
    The color code represents the size of the local magnetic moments with up (red) and down (blue) polarization. }
    \label{fig:EkDOS_magnetic}
\end{figure} 

\begin{figure}[htbp]
    \centering
    \includegraphics[width=0.48\textwidth]{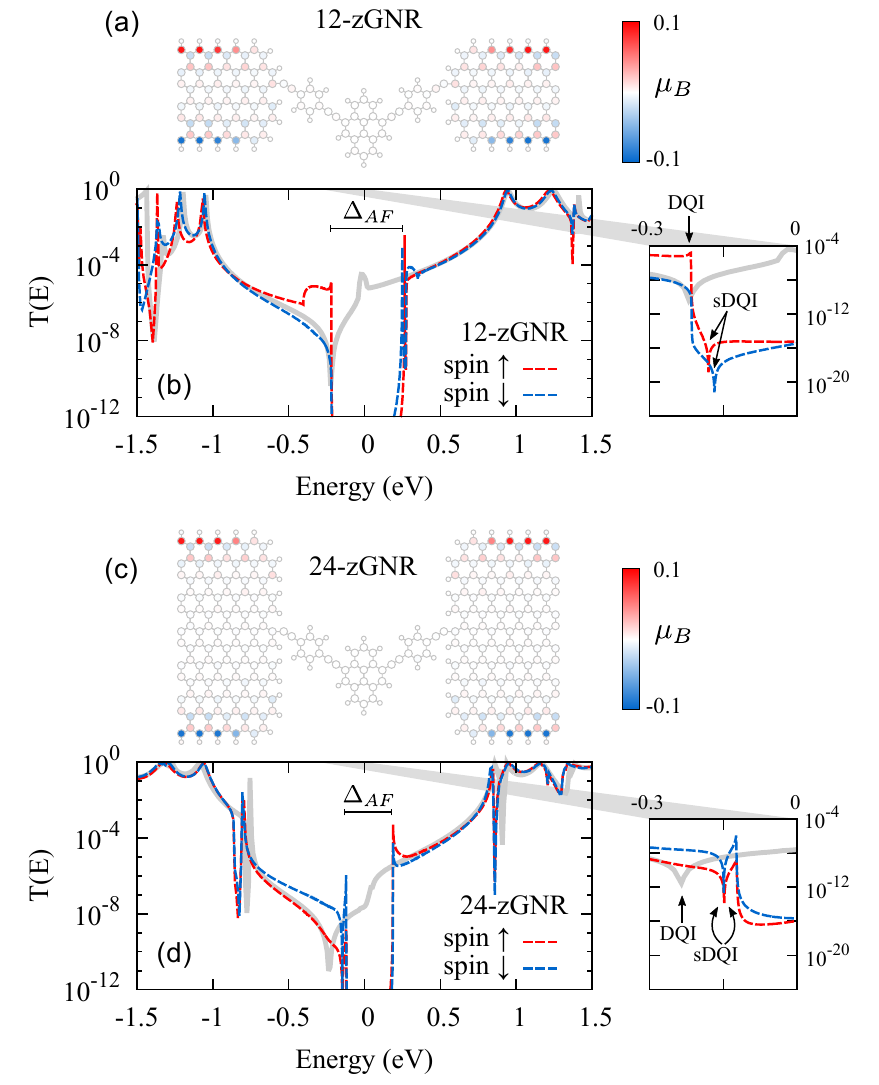}
    \caption{(a,c) Spin-ordered structures with the pyrene molecule in the meta configuration 
    and magnetic zGNR electrodes. 
    The color code represents the size of the local magnetic moments with up (red) and down (blue) polarization. 
    (b,d) Spin-resolved transmission functions (red and blue dashed lines) 
    compared to those of spin-paired calculations (grey solid lines). 
    The presence of the AF gap $\Delta_{AF}$ results in a strong suppression of the transmission function
    close to the Fermi level, as also highlighted in each inset. 
    See text for discussion. }
    \label{fig:GNR_Te_magnetic}
\end{figure}

For the sake of completeness, in Fig.~\ref{fig:EkDOS_magnetic} we show the bandstructure of periodic 12-zGNR 
and the spatial distribution of the magnetic moments in the N\'{e}el AF state. 
Note that the bandstructures for the two spin states are degenerate in all bands.~\cite{son2006energy} 
The characteristic flat bands at the Fermi level, observed in Fig.~\ref{fig:EkDOS}(b), 
disappear in favor of an AF gap $\Delta_{AF}$. 
The band splitting at the $Z$ point of the one-dimensional Brillouin zone 
(i.e., $k=\pi/a$, with $a$ the lattice constant) is approximately $0.5$~eV and weakly width-dependent, 
while the narrower splitting along the $\Gamma-Z$ path is inversely proportional to the zGNR width, 
in agreement with the literature.~\cite{son2006energy} 

In order to evaluate the transport properties and the interplay between DQI and magnetism, 
we perform spin-polarized calculations for both the leads and the scattering region 
with the pyrene molecule contacted in the meta configuration. 
The structures considered and the corresponding spin-polarized transmission functions 
are shown in Fig.~\ref{fig:GNR_Te_magnetic}. 
As expected, the transmission functions in Figs.~\ref{fig:GNR_Te_magnetic}(b,d) 
display a sizable AF gap at the Fermi level, which depends on the width of the zGNR.~\cite{son2006energy}  
Weak spin-polarization effects are observed close to the Fermi level, 
where the transport mechanisms involves the states localized at the ZZ egdes. 
Instead, the features attributed to MOs with a strong pyrene character barely differ from those 
observed in the spin-paired calculations, as the molecule remains non-magnetic. 
The transmission functions display weakly spin-dependent DQI (sDQI) antiresonances, 
appearing either inside (for 12-zGNRs) or outside (for 24-zGNRs) the AF gap, 
as clearly shown in the corresponding insets. 
It is interesting to note that DQI results in a suppression of the transmission even inside the magnetic gap, 
since DQI is associated to an actual zero of the Green's function, 
rather than just a suppression of spectral weight.~\cite{pedersen2014quantum,valli2018quantum,valli2019interplay} 
In the limit of $\infty$-zGNR, magnetism is expected to become irrelevant, 
since the ZZ edges and hence the associated sublattice imbalance disappear.

From our analysis, we conclude that the existence of DQI induced antiresonances 
is not directly affected by the edge magnetism on the graphene electrodes. 
However, the possibility of identifying DQI unambiguously remains conditional to 
the details of the magnetic order, which is controlled by the width and the topology of the GNR electrodes. 
Let us also note that in the cases discussed above, the position of the sDQI antiresonances 
is almost independent on the spin polarization, since the pyrene core is not magnetic. 
Instead, if the source of QI itself is magnetic, one can expect also 
a spin-dependent splitting of the DQI energy, $\epsilon^{\textrm{DQI}}_{\sigma}$,  
as reported in recent studies.~\cite{valli2018quantum,valli2019interplay,phung2020spin}

\section*{Conclusions and outlook}\label{sec:outlook} 

We presented a thorough analysis of the effects of different anchoring groups and electrodes 
on the DQI properties in pyrene based single-molecule junctions. 
Simple theoretical approaches allow to reliably predict the occurrence or absence of DQI 
in specific contact configurations because it stems from molecular symmetry. 
However, the role of the chemical nature of all components of the junction 
for the observability of DQI features can only be unraveled with DFT+NEGF calculations. 

In our study, we find that the linkers play an important role with Au electrodes, 
as the charge redistribution within the junction determines the relative energy position 
of the antiresonance with respect to the Fermi level and the frontier MOs. 
This eventually determines to which extent DQI affects the current-voltage characteristics of the junction. 
For pyrene junctions, the choice of Au electrodes 
allows to clearly observe DQI in the electronic transmission function,  
as the electrode do not introduce additional transport features. 

In the case of graphene, the electronic properties at the Fermi level 
are strongly dependent on the width of the electrodes and the topology of the edges. 
As a consequence, the electrodes can modulate the transport properties of the junction 
and introduce transmission oscillations close to the Fermi level, 
which may obscure or even resemble QI antiresonances.~\cite{gehring2017distinguish} 
We were able to disentangle the features arising from the molecule and those arising from the electrodes, 
and we traced back the origin of each transmission feature to the topology of the edges. 
In this respect, we provide clear indications to guide the interpretation of experimental results, 
which is generally difficult, especially when the energy scales of the molecule  
and that of the electrodes become comparable. 
We also stress that one of the advantages of graphene over metal electrodes 
is a better control on the position of the Fermi level. 
This can be achieved by gating, which allows to maximize the effect of DQI 
(or of the edge states for that matter) on the current-voltage characteristics 
and is therefore of pivotal importance for technological applications exploiting QI. 

Finally, we considered the effects of magnetic order in graphene electrodes, 
which arises due to the local sublattice imbalance at the ZZ edges. 
Since magnetism originates at the edge states, it mainly affects 
the electronic and transport properties in an energy window around the Fermi level. 
For zGNRs, a transmission gap of magnetic origin in the electrodes 
can potentially prevent a clear observation of DQI. 
For the junctions we considered, the AF spin-ordered pattern resulting in a global singlet state, 
and the transmission displays a negligible spin polarization. 
In the presence of a net magnetic moment one would expect a significant 
spin splitting of the transmission of the majority and minority spin channels. 

While the present analysis revolves around prototypical single-molecule junction setups, 
it also poses the basis for future studies. 
When the focus shifts to complex graphene-based devices, 
such as nanoconstrictions~\cite{gehring2016quantum} 
or junctions with graphene quantum dots,~\cite{valli2018quantum,valli2019interplay,nictua2020robust} 
the source QI may differ in the presence of the electrodes with multiple contacts.
The additional contacts increase the number of potential transmission channels 
and QI patterns can emerge within individual or between different channels, thereby
increasing the complexity of the analysis. 
Junctions with multiple anchors can display either a low- or a high- conductance, 
depending on the connectivity of the molecule,~\cite{markin2020conductance} 
which can be electrochemically controlled.~\cite{darwish2012observation,grunder2007cruciform} 
A systematic investigation of multiple-contact junctions 
can open new avenues to exploit QI effects in the framework of graphene nanoelectronics.

\section*{Conflicts of interest}
There are no conflicts to declare.

\section*{Acknowledgments}
We thank G.~Gandus, G.~Kastlunger, and J.~V\"olkle for helpful discussions. 
We acknowledges financial support from the Austrian Science Fund (FWF) through project P~31631. 
Calculations have been performed on the Vienna Scientific Cluster (VSC), project No.~71279.

\bibliographystyle{apsrev}
\bibliography{arXiv}

\end{document}